\documentclass[structabstract]{aa}  
\usepackage{graphicx}
\usepackage{txfonts}
\usepackage{natbib}
\begin{document}
\title{The onset of high-mass star formation in the direct vicinity of
  the galactic mini-starburst W43\thanks{The Herschel, APEX, Nobeyama
    and PdBI data are available in electronic form at the CDS via
    anonymous ftp to cdsarc.u-strasbg.fr (130.79.128.5) or via
    http://cdsweb.u-strasbg.fr/cgi-bin/qcat?J/A+A/}}


   \author{H.~Beuther
          \inst{1}
          \and
          J.~Tackenberg
           \inst{1}
          \and
          H.~Linz
          \inst{1}
          \and
          Th.~Henning
           \inst{1}
          \and
          O.~Krause
          \inst{1}
          \and
          S.~Ragan
          \inst{1}
          \and
          M.~Nielbock
          \inst{1}
          \and
          R.~Launhardt
          \inst{1}
          \and
          A.Schmiedeke
          \inst{1,3}
          \and
          F.~Schuller
          \inst{2}
          \and
          P.~Carlhoff
          \inst{3}
          \and
          Q.~Nguyen-Luong
          \inst{4}
          \and
          T.~Sakai
          \inst{5}
           }
   \institute{$^1$ Max-Planck-Institute for Astronomy, K\"onigstuhl 17,
              69117 Heidelberg, Germany, \email{name@mpia.de}\\
$^2$ Max Planck Institute for Radioastronomy, Auf dem H\"ugel 69,
              53121 Bonn, Germany\\
              $^3$ University of Cologne, Z\"ulpicher Strasse 77, 50937 K\"oln, Germany\\
$^4$ Laboratoire AIM Paris-Saclay, CEA/IRFU, CNRS/INSU, Université Paris Diderot, Service d'Astrophysique, Bât. 709, CEA-Saclay, 91191, Gif-sur-Yvette Cedex, France\\
$^5$ Institute of Astronomy, The University of Tokyo, Osawa, Mitaka, Tokyo 181-0015, Japan}



\abstract
{The earliest stages of high-mass star formation are still poorly
  characterized. Densities, temperatures and kinematics are crucial
  parameters for simulations of high-mass star formation. It is
  also unknown whether the initial conditions vary with environment.}
{We want to investigate the youngest massive gas clumps in the environment of
  extremely active star formation. }
{We selected the IRDC\,18454 complex, directly associated with the W43
  Galactic mini-starburst, and observed it in the continuum emission
  between 70\,$\mu$m and 1.2\,mm with Herschel, APEX and the 30\,m
  telescope, and in spectral line emission of N$_2$H$^+$ and $^{13}$CO
  with the Nobeyama 45\,m, the IRAM 30\,m and the Plateau de Bute
  Interferometer.}
{The multi-wavelength continuum study allows us to identify clumps
  that are infrared dark even at 70\,$\mu$m and hence the best
  candidates to be genuine high-mass starless gas clumps. The spectral
  energy distributions reveal elevated temperatures and luminosities
  compared to more quiescent environments. Furthermore, we identify a
  temperature gradient from the W43 mini-starburst toward the starless
  clumps. We discuss whether the radiation impact of the nearby
  mini-starburst changes the fragmentation properties of the gas
  clumps and by that maybe favors more high-mass star formation in
  such an environment. The spectral line data reveal two different
  velocity components of the gas at 100 and 50\,km\,s$^{-1}$. While
  chance projection is a possibility to explain these components, the
  projected associations of the emission sources as well as the
  prominent location at the Galactic bar -- spiral arm interface also
  allow the possibility that these two components may be spatially
  associated and even interacting.}
{High-mass starless gas clumps can exist in the close environment of
  very active star formation without being destroyed. The impact of
  the active star formation sites may even allow for more high-mass
  stars to form in these 2nd generation gas clumps. This particular
  region near the Galactic bar -- spiral arm interface has a broad
  distribution of gas velocities, and cloud interactions may be
  possible.}  
   \maketitle

\section{Introduction}
\label{intro}

The initial conditions required to form massive stars is one of the
major topics in high-mass star formation research today. The advent of
new observational capabilities often opened the window to earlier and
earlier evolutionary stages in that field. While the IRAS satellite
combined with cm continuum surveys of the galactic plane revealed the
populations of ultracompact H{\sc ii} regions and high-mass
protostellar objects (HMPOs, e.g.,
\citealt{wc89,wc1989b,kurtz1994,molinari1996,sridha}), the advent of
the mid-infrared satellites ISO, MSX and Spitzer allowed to access
even earlier evolutionary stages, namely the infrared dark clouds
(IRDCs, e.g.,
\citealt{perault1996,egan1998,carey2000,sridharan2005,simon2006,peretto2009}).
However, it turned out that most of the so far studied IRDCs host weak
24\,$\mu$m emission sources and drive molecular outflows, both strong
indicators for star formation activity (e.g.,
\citealt{beuther2005d,beuther2007a,beuther2007g,rathborne2006,motte2007,cyganowski2008}).
Therefore, we are still lacking a thorough understanding of the
initial conditions for high-mass star formation prior to any star
formation activity. The advent of the Herschel far-infrared satellite
\citep{A&ASpecialIssue-HERSCHEL} now offers the unique opportunity to
identify targets for these initial conditions and to study their
physical and chemical properties in detail. Major characteristics of
these earliest evolutionary stages are, that the massive gas cores
with masses in excess of several 100\,M$_{\odot}$ observed at (sub)mm
wavelengths are still dark at 70\,$\mu$m, hence they have not formed
any protostellar objects to Herschel sensitivity limits yet.

The Herschel guaranteed time key project EPOS (Early Stages of Star
Formation, P.I.~O.~Krause) is a dedicated observation campaign
targeting 45 IRDCs that are promising candidates to search for and
to study these initial conditions. Early results from the first data
revealed already exciting results, e.g., we identified candidate
high-mass starless cores (HMSCs) as well as very weak 70\,$\mu$m
sources within IRDCs that may be the very first manifestation of
high-mass star formation \citep{beuther2010b,henning2010,linz2010}.

\begin{figure*}[htb]
\includegraphics[width=0.99\textwidth]{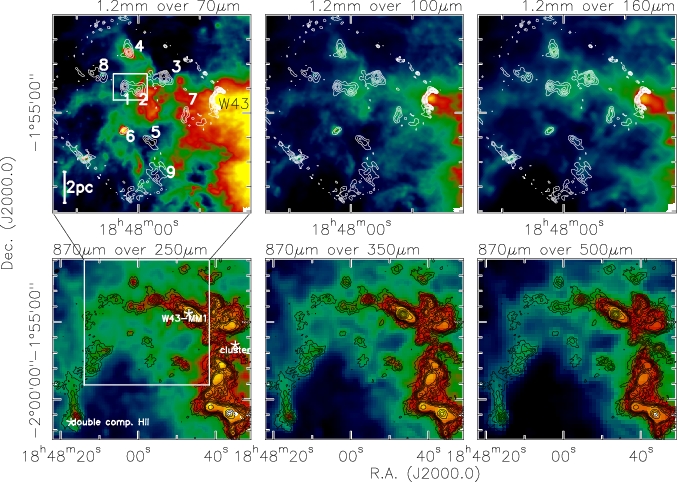}
\caption{Compilation of the continuum data from 70\,$\mu$m to 1.2\,mm
  wavelength as labeled in each panel. The top-row presents the data
  on a smaller spatial scale as marked in the bottom-left panel. The
  color-scale is chosen in each image individually to highlight the
  most important features. Contour levels of the 870\,$\mu$m data
  start at the 3$\sigma$ levels of 0.27\,mJy\,beam$^{-1}$ and continue
  in in 3$\sigma$ steps to 2.7\,Jy\,beam$^{-1}$, from where they
  continue in 2.7\,Jy\,beam$^{-1}$ steps. The 1.2\,mm data are
  contoured in 3$\sigma$ levels of 30\,mJy\,beam$^{-1}$. The numbers
  in the top-left panel label the sub-cores, the white small box there
  marks the region shown in Fig.~\ref{18454_spitzer}, and a scale-bar
  for a distance of 5.5\,kpc is shown as well. The crosses in
  the bottom-left panel mark the positions of a double-component radio
  recombination line H{\sc ii} region from \citet{anderson2011}, the
  W43-MM1 position from \citet{motte2003} as well as the approximate
  center of the Wolf-Rayet/OB cluster \citep{blum1999,motte2003}.}
\label{spire_pacs}
\end{figure*}

\paragraph{The IRDC\,18454 close to W43:} One particularly interesting
region within the sample are the IRDCs associated with the IRAS source
IRAS\,18454-0158. Henceforth, we will call the region IRDC\,18454.
While the IRAS source was first studied in the framework of the HMPO
survey by \citet{sridha} and \citet{beuther2002a}, it was soon
recognized that several of the sub-sources are indeed infrared dark
\citep{sridharan2005}. Another curiosity of that region is that it
harbors two distinctly different velocity components, one around 50
and one around 100\,km\,s$^{-1}$ \citep{sridharan2005,beuther2007g}.
It is also interesting to note that only the 100\,km\,s$^{-1}$
component shows SiO emission, indicative of a molecular outflow,
whereas we do not detect any SiO from the 50\,km\,s$^{-1}$ component
\citep{beuther2007g}. Furthermore, the measured H$^{13}$CO$^+$(1--0)
line-width from the 100\,km\,s$^{-1}$ component is also broader than
that at 50\,km\,s$^{-1}$ (2.7 versus 1.7\,km\,s$^{-1}$,
\citealt{beuther2007g}), also indicating that the 50\,km\,s$^{-1}$
component is in a less turbulent state. The kinematic distances
derived by \citet{beuther2007g} for the 50 and 100\,km\,s$^{-1}$
components, using the \citet{brand1993} rotation curve were 3.5 and
6.4\,kpc, respectively. Applying the new rotation curve by
\citet{reid2009}, we now get kinematic distances of 3.3 and 5.5\,kpc
for both components, respectively. The 5.5\,kpc distance corresponds
well to the distance derived for the large H{\sc ii} region W43 by
\citet{wilson1970}. A similar distance of $\sim 6$\,kpc was also
recently derived for the nearby red supergiant cluster RSGC3
\citep{negueruela2011}). We will discuss later whether these are
really two different components just chance-projected together on the
plane of the sky, or whether it may be interacting cloud components at
similar distances.

In the direct vicinity of that region we find the large Galactic
mini-starburst region W43 (Fig.~\ref{spire_pacs}), which was already
studied in detail in the past with a broad wavelength coverage (e.g.,
\citealt{smith1978,lester1985,blum1999,motte2003,nguyen2011})
including also early Herschel results \citep{bally2010,elia2010}. The
IRDC\,18454 is located at the northeastern edge of the Z-shaped
filament first discussed in \citet{motte2003} (see also
Fig.~\ref{spire_pacs}). This region is very luminous with $L\sim
3\times 10^6$\,L$_{\odot}$, and it contains sources at different
evolutionary stages from a Wolf-Rayet cluster to active young
star-forming cores (e.g., \citealt{blum1999,motte2003}).
\citet{nguyen2011} recently discussed several broad velocity
components in that regions and suggested that this whole complex might
have formed via converging gas flows due to its special location at
the interface of the Galactic bar with the inner Scutum spiral arm
\citep{benjamin2005,lopez2007,rodriguez2008}. Different velocity
components were also identified in atomic HI emission and absorption
\citep{liszt1993}. However, our target region IRDC\,18454 which hosts
supposedly the earliest evolutionary stages, has not been subject to a
detailed investigation yet.

Among others, we like to address the following questions: Do we
identify bona-fide high-mass starless cores at all? If yes, what are
their physical properties, e.g., their temperatures and turbulent
line-widths? Are the two velocity components spatially associated or
are they rather chance alignments along the line of sight? Does the
mini-starburst W43 impose any impact on the earliest evolutionary
stages or are the young cores evolving independently of that? Do we
find any signature of triggering?

\section{Observations and archival data} 
\label{obs}

\subsection{Herschel}

\begin{figure*}[htb]
\includegraphics[width=0.99\textwidth]{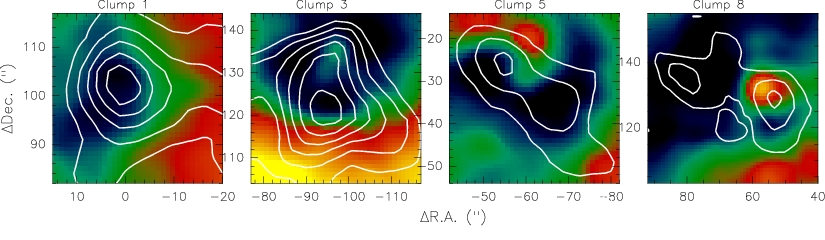}
\caption{Zoom into the 70\,$\mu$m dark cores with 1.2\,mm continuum
  contours on the 70\,$\mu$m color-scale emission. The color-scale is
  again chosen to highlight the most important features. The 1.2\,mm
  data are contoured in 3$\sigma$ levels of 30\,mJy\,beam$^{-1}$. The
  0 position is at R.A.~(J2000.0)18:48:02.160 and Dec.~(J2000.0)
  -01:55:38.0.}
\label{zooms}
\end{figure*}

The cloud complex with a size of $\sim6'\times 6'$ was observed with
PACS \citep{A&ASpecialIssue-PACS} on Herschel
\citep{A&ASpecialIssue-HERSCHEL} on 2010 March 9.  Scan maps in two
orthogonal directions with scan leg lengths of $18'$ and $6'$,
respectively, were obtained with the medium scan speed of $20''$/s.
The raw data have been reduced up to level-1 with the HIPE software
\citep{A&ASpecialIssue-PACS,ott2010}, version 6.0, build 1932. Besides the
standard steps, we applied a 2$^{nd}$level deglitching, in order to
remove bad data values from a given pixel map by $\sigma$-clipping the
flux values which contribute to each pixel. We used the time-ordered
option and applied a 25$\sigma$ threshold. The final level-2 maps were
processed using Scanamorphos version 8 (Roussel 2011, subm.). Since
the field of view contains bright emission on scales larger than the
map, we applied the ``galactic'' option and included the
non-zero-acceleration telescope turn-around data.  The flux correction
factors provided by the PACS ICC team were applied.

Maps at 250, 350, and 500\,$\mu$m were obtained with SPIRE
\citep{A&ASpecialIssue-SPIRE} on 2010 March 11. Two $14'$ scan legs were used to cover
the source. The data were processed up to level-1 within HIPE,
developer build 5.0, branch 1892, calibration tree 5.1 using the
standard photometer script (POF5\_pipeline.py, dated 2.3.2010)
provided by the SPIRE ICC team.  The resulting level-1 maps have been
further reduced using Scanamorphos, version 9 (patched, dated
08.03.2011). This version included again the essential de-striping for
maps with less than 3 scan legs per scan. In addition, we used the
``galactic'' option and included the non-zero-acceleration
telescope turn-around data. Color corrections were applied according
to the SPIRE Manual. 

\subsection{Plateau de Bure Interferometer}

We observed IRDC\,18454-1 with the Plateau de Bure Interferometer
during five nights in October and November 2009 at 93\,GHz in the C
and D configurations covering projected baselines between
approximately 13 and 175\,m.  The 3\,mm receivers were tuned to
92.835\,GHz in the lower sideband covering the N$_2$H$^+$(1--0) as
well as the 3.23\,mm continuum emission.  For continuum measurements
we placed six 320\,MHz correlator units in the band, the spectral
lines were excluded in averaging the units to produce the final
continuum image.  Temporal fluctuations of amplitude and phase were
calibrated with frequent observations of the quasars 1827+062 and
1829-106.  The amplitude scale was derived from measurements of MWC349
and 3c454.3. We estimate the final flux accuracy to be correct to
within $\sim 15\%$. The phase reference center is R.A.~(J2000.0)
18:48:02.21 and Dec.~(J2000.0) -01:53:55.79, and the velocity of rest
$v_{\rm{lsr}}$ is $\sim$51.9 and 99.6\,km\,$s^{-1}$ for the two
velocity components, respectively. The synthesized beam of the line
and continuum data is $3.9''\times 3.2''$ with a P.A. of 34 degrees.
The $3\sigma$ continuum rms is 0.27\,mJy\,beam$^{-1}$. The $3\sigma$
rms of the N$_2$H$^+$(1--0) data measured from an emission-free
channel with a spectral resolution of 0.2\,km\,s$^{-1}$ is
21\,mJy\,beam$^{-1}$.

\subsection{Nobeyama 45\,m telescope}

The N$_2$H$^+$ data has been observed using the BEARS receiver at the
NRO 45\,m telescope in Nobeyama, Japan. The receiver has been tuned to
$93.17346$\,MHz, covering the full hyperfine structure of the
N$_2$H$^+(1-0)$ transition. At this frequency the telescope beam is
$18.2''$ and the observing mode provides a spectral resolution of
$0.2$\,km\,s$^{-1}$ at a bandwidth of 32\,MHz. The different velocity
components have been observed one in April 2010 with an average system
temperature of $T_{\rm{sys}}=206$\,K, the second in June, at slightly
lower $T_{\rm{sys}}$ . The software package nostar \citep{sawada2008}
was used for the data reduction, sampling the data to a pixel size of
$7''$ with a spheroidal convolution and smoothing the spectral
resolution to $0.5$\,km\,s$^{-1}$. The rms of the final map is $\sim
0.125$\,K and $\sim 0.09$\,K for the 100\,km\,s$^{-1}$ and
50\,km\,s$^{-1}$ map, respectively. Strong winds hampered the pointing
for part of the observations, contributing to spatial uncertainties.

\subsection{Archival data from Spitzer, APEX and the IRAM 30\,m}

The MIPS 24\,$\mu$m data (from MIPSGAL, \citealt{carey2009}) as well
as the IRAC 8\,$\mu$m observations (from GLIMPSE,
\citealt{churchwell2009}) are taken from the Spitzer archive. The
1.2\,mm continuum data were first presented in \citet{beuther2002a}
and the APEX 870\,$\mu$m data are part of the ATLASGAL survey of the
Galactic plane \citep{schuller2009}. Beam sizes of the two datasets are
$10.5''$ and $19.2''$, respectively. The $1\sigma$ rms values are 10 and
90\,mJy\,beam$^{-1}$, respectively.

\section{Results}

\subsection{Continuum emission}
\label{continuum}

\begin{figure*}[htb]
\includegraphics[width=0.99\textwidth]{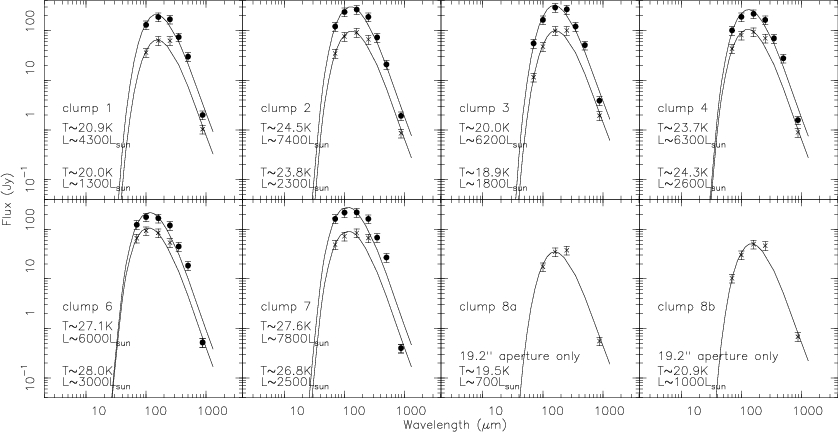}
\caption{Spectral energy distributions (SEDs) of the discussed clumps.
  The fluxes are extracted for apertures of $36.6''$ and $19.2''$. For
  clumps 8a and 8b only the smaller apertures of $19.3''$ were used.
  Error bars represent 20\% uncertainties. Each panel lists the
  corresponding temperatures $T$ and bolometric luminosities $L$ for
  both fits top and bottom, respectively. Clumps 1, 3 and 8a are
  starless clump candidates.}
\label{seds}
\end{figure*}

Figure \ref{spire_pacs} gives an overview of the target regions in all
accessible continuum bands from the far-infrared at 70\,$\mu$m to the
1.2\,mm band. The bottom panel of Fig.~\ref{spire_pacs} shows the
large-scale emission covering prominently the Z-shaped strong submm
emission from the mini-starburst region W43 in the west of the images.
The morphology of all images longward of 250\,$\mu$m wavelengths
agrees well in all bands. While these data would also allow us an
analysis of the W43 region itself (see also \citealt{bally2010}), for
this project we concentrate on the earliest evolutionary stages in the
north-eastern part of the region highlighted in the top panels of
Fig.~\ref{spire_pacs}. The high spatial resolution of the Herschel
PACS images between 70 and 160\,$\mu$m (between $\sim$5.6$''$,
$\sim$6.8$''$ and $\sim$11.4$''$) and the 1.2\,mm map from the IRAM
30\,m ($\sim$10.5$''$) allow us to spatially differentiate different
components. While the long-wavelength data from 250\,$\mu$m onwards
are structurally similar, a spatial comparison of the 70\,$\mu$m with
the 1.2\,mm data shows significant differences. In simple terms, the
emission longward of 250\,$\mu$m is largely dominated by the cold dust
from the natal gas and dust clump\footnote{In the following we refer
  to clumps for scales covered by the single-dish observations
  exceeding 0.25\,pc (or 50000\,AU) at the assumed distance of
  5.5\,kpc , whereas the smaller scales traced by the Plateau de Bure
  observations (on the order of 20000\,AU) are refered to as cores
  \citep{williams2000,beuther2006b,bergin2007}.}, whereas the emission
shortward of 100\,$\mu$m becomes dominated by warmer dust components,
which may be heated by internal early star formation processes. Some
of the mm clumps exhibit strong mid-infrared emission whereas others
are either weak or show no emission at all at 70\,$\mu$m.  This is in
contrast to the neighboring W43 region, which is strong in all
wavelengths covered here. This already outlines the different
evolutionary stages of the main W43 region and the infrared dark gas
clumps we are investigating here.

\begin{table}[htb]
\caption{Clump parameters from the 1.2\,mm data}
\label{parameters}
\begin{tabular}{lrrrrrr}
\hline \hline
\# & R.A. & Dec. & $S_{\rm{peak}}$ & $S$ & N$_{\rm{H}_2}$ & $M$ \\
   & (J2000.0) & (J2000.0) & $\left(\frac{\rm{mJy}}{\rm{beam}}\right)$ & (mJy) & $\left(\frac{10^{23}}{\rm{cm}^2}\right)$ & (M$_{\odot}$) \\
\hline
1 & 18:48:02.23 & -01:53:56.4 & 178 & 479 & 1.6 & 609  \\
2 & 18:47:59.99 & -01:54:07.5 & 157 & 468 & 1.4 & 595  \\
3 & 18:47:55.86 & -01:53:36.8 & 203 & 816 & 1.8 & 1037 \\
4 & 18:48:01.56 & -01:52:30.9 & 175 & 491 & 1.6 & 624  \\
5a& 18:47:58.50 & -01:56:03.9 & 102 & 125 & 0.9 & 159  \\
5b& 18:47:57.68 & -01:56:13.9 &  88 & 169 & 0.8 & 215  \\
6 & 18:48:02.28 & -01:55:43.2 &  87 & 108 & 0.8 & 137  \\
7 & 18:47:52.21 & -01:54:54.5 & 100 & 123 & 0.9 & 156  \\
8a& 18:48:07.60 & -01:53:21.6 &  81 & 108 & 0.7 & 137  \\
8b& 18:48:05.74 & -01:53:28.5 & 103 & 178 & 0.9 & 226  \\
\hline \hline
\end{tabular}
\end{table}

Considering an evolutionary sequence, the youngest sources without
internal heating sources should be dark even at 70\,$\mu$m wavelength.
As soon as a central core/protostar starts producing its own
luminosity either by accretion shocks or nuclear synthesis, emission at
70\,$\mu$m, 24\,$\mu$m and at shorter wavelengths will become
detectable (e.g., \citealt{beuther2010b,henning2010,motte2010}).  To
examine the youngest starless candidate clumps in more detail, Figure
\ref{zooms} presents zoomed in images of the selected mm clumps that
show no or only weak 70\,$\mu$m emission. The 1.2\,mm emission peaks
of clumps 1, 3, 5a, 8a and 9 are all non-detections at 70\,$\mu$m
wavelengths.  However, there are subtle differences between the
clumps. For example, clump 1 is a clear 70\,$\mu$m absorption feature
whereas clump 3 shows 70\,$\mu$m emission associated with the
north-eastern extension of the 1.2\,mm cold dust emission. While this
indicates that clump 1 is likely still in a starless evolutionary
phase, star formation may have already started in clump 3, although
the column density peak still appears devoid of active star formation
signatures.

\begin{figure*}[tb]
\includegraphics[width=0.46\textwidth]{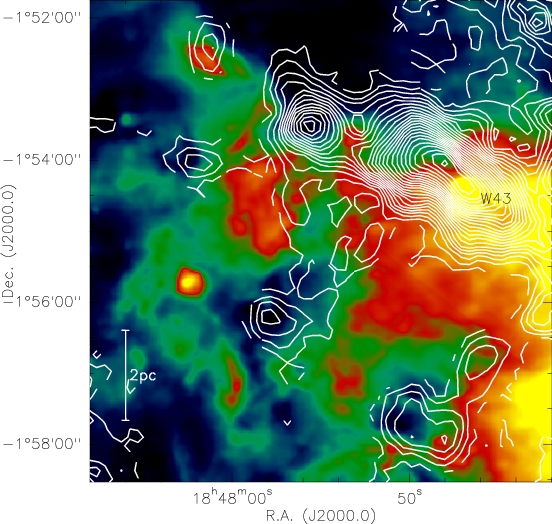}
\includegraphics[width=0.52\textwidth]{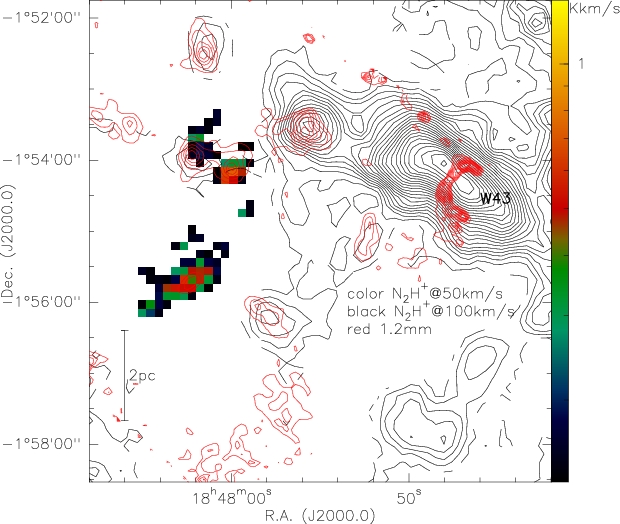}
\caption{{\bf Left:} N$_2$H$^+$(1--0) contours over 70\,$\mu$m
  emission. The color-scale is again chosen to highlight the most
  important features. The N$_2$H$^+$ emission is integrated from 91 to
  106\,km\,s$^{-1}$, and contours start at 0.6 and continue in
  1\,K\,km\,s$^{-1}$ levels. {\bf Right:} N$_2$H$^+$(1--0) and 1.2\,mm
  emission. The color-scale shows the N$_2$H$^+$ emission integrated
  from 49 to 53\,km\,s$^{-1}$ and the black contours present the
  N$_2$H$^+$ emission integrated from 91 to 106\,km\,s$^{-1}$. The
  contours start at 0.6 and continue in 1\,K\,km\,s$^{-1}$ levels. The
  red 1.2\,mm emission is contoured in 3$\sigma$ levels of
  30\,mJy\,beam$^{-1}$. Scale-bars are shown at the bottom-right of
  each panel.}
\label{n2hplus}
\end{figure*}

Assuming optically this dust continuum emission at (sub)mm wavelengths
at a typical temperature of 20\,K (see section \ref{sedsection}) with
a dust opacity index $\beta =2$ (corresponding to
$\kappa_{\rm{1.2mm}}\sim 0.4$\,cm$^2$g$^{-1}$ representing the general
ISM, e.g., \citealt{mathis1977,ossenkopf1994}), we can calculate the
clump masses and peak column densities following the classical
approach by \citet{hildebrand1983}. An adaptation of the equations can
be found in \citet{beuther2002a,beuther2002erratum}. The resulting
values derived from the 1.2\,mm data are presented in Table
\ref{parameters}.  Although the derived masses are uncertain by about
a factor 5 based on mainly temperature, $\beta$ and gas-to-dust mass
ratio uncertainties (e.g.,
\citealt{ossenkopf1994,beuther2002a,draine2007}), nevertheless, the
average clump masses are relatively high. If one assumes an initial
mass function (IMF) following \citet{kroupa2001} with an approximate
star formation efficiency of 30\%, the initial gas clumps should have
masses in excess of 1000\,M$_{\odot}$ in order to be capable to form
at least one 20\,M$_{\odot}$ star. In this picture, at least clump 1
can be considered as one of the best candidates of being a high-mass
starless clump. Clump 3 also full-fills the mass criteria, but it
exhibits two point sources at the edge of the clump, hence some star
formation activity may have just started. Nevertheless, clump 3 should
also be in a very young evolutionary stage and resemble the initial
conditions of high-mass star formation well.

The estimated column densities based on the 30\,m single dish data are
in the regime $0.7-1.8\times 10^{23}$\,cm$^{-2}$ (Table
\ref{parameters}).  While this is slightly below the typically
discussed threshold of $\sim$1\,g\,cm$^{-2}$ (e.g., \citealt{krumholz2008b},
corresponding to $\sim$3$\times 10^{23}$\,cm$^{-2}$) this is certainly
no counterargument since these are average values of the beam size of
$10.5''$ that corresponds to linear spatial scales of approximately
58000\,AU or a quarter of a pc. Higher resolution observations will
reveal significantly larger column densities (section \ref{pdbi}).
Therefore, also from that point of view, the regions qualify as likely
extremely young high-mass star-forming regions.

\subsection{Spectral energy distributions (SEDs)}
\label{sedsection}

The Herschel data allow us to extract SEDs for the gas and dust
clumps, and by that to determine the bolometric luminosities and
temperatures. Since we are targeting very young regions that are not
detected as point sources in the Herschel bands, classical
PSF-photometry or point source flux extraction where the background is
subtracted is difficult. Furthermore, we like to include all Herschel
bands for our SED fits. Therefore, we extract the peak fluxes at all
wavelengths from maps smoothed to the $36.6''$ beam of the 500\,$\mu$m
data.  Furthermore, we also extract the peak fluxes from maps smoothed
to the $19.2''$ resolution of the 870\,$\mu$m ATLASGAL data, ignoring
the 350 and 500\,$\mu$m maps then. Since clumps 8a and 8b are too
close to be separated by the $36.6''$ beam, for these clumps we only
use the $19.2''$ maps. For clumps 5a and 5b, where even at
250\,$\mu$m the two peaks merge into one, no proper flux determination
was possible. At the assumed distance of 5.5\,kpc, the apertures of
$36.6''$ and $19.2''$ correspond to linear apertures of $\sim 1$\,pc
and $\sim 0.5$\,pc, respectively.

One inherent problem in flux measurements from Herschel data is that
the diffuse Galactic background level is unknown and therefore hard to
correct for. In particular in a region like the one presented here
where at all wavelengths significant emission is seen throughout the
whole maps. One should note that the problem is more severe when
targeting very young or starless regions that do not exhibit clear
point sources, because for more evolved point-source like structure,
proper PSF-photometry better allows to subtract a background level.
To approximate a background level, we define polygons at the map edges
with low emission regions. The mean of these regions is then
subtracted from our measured fluxes.

\begin{figure*}[htb]
\includegraphics[width=0.99\textwidth]{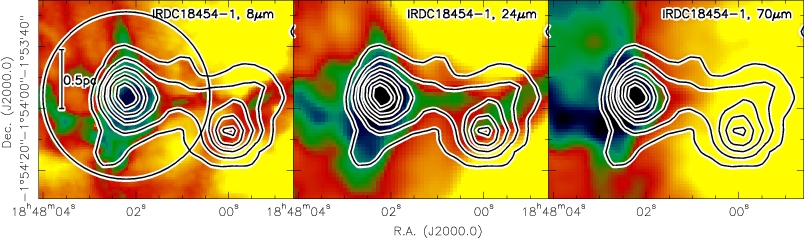}
\caption{Zoom into the two central clumps 1 and 2 (see
  Fig.~\ref{spire_pacs}). The left and middle panels present in
  color-scale the Spitzer 8 and 24\,$\mu$m emission, whereas the right
  panel shows the Herschel 70\,$\mu$m image. Color-scale is again to
  highlight the main features with dark low and yellow high
  intensities. The contours show the 1.2\,mm continuum emission
  contoured in 3$\sigma$ levels of 30\,mJy\,beam$^{-1}$. The circle
  outlines the FWHM of the primary beam of the PdBI observations, and
  a scale-bar is shown in the left panel as well.}
\label{18454_spitzer}
\end{figure*}

The SEDs were fitted using modified blackbody functions accounting for
the wavelength-dependent emissivity of the dust. The assumed dust
composition follows \citet{ossenkopf1994} with thin ice mantles, and
the assumed gas-to-dust mass ratio is 100. Figure \ref{seds} presents
the SEDs of all clumps (except 5a and 5b) as derived for the apertures
of $36.6''$ and $19.2''$. While the overall spread of fitted dust
temperatures is relatively low (between $\sim$18 and $\sim$26\,K), we
identify on average lower temperatures toward the mid-infrared dark
clumps 1, 3 and 8 (19 to 20\,K) compared to the clumps with embedded
objects (21 to 28\,K).

We compared the temperatures derived from our SED fitting (here and
section \ref{luminous}) with the temperature map recently produced by
\citet{battersby2011}. The main differences between the two datasets
is that Battersby and collaborators used the HIGAL data
\citep{molinari2010}, and they were able to subtract a background in a
systematic fashion because of the much larger spatial coverage.  The
temperature comparison shows that in the low-$T$ regime ($\leq 25$\,K)
the temperatures agree very well, whereas in the high-$T$ regime
($\geq 30$\,K) our analysis underestimates their temperatures.  Since
here we are mainly interested in the cold gas clumps, our approach is
perfectly justifiable.

The other important parameter one can derive from the SEDs is the
bolometric luminosities of the regions. As expected, for smaller
apertures one finds lower bolometric luminosities. Nevertheless, it is
interesting to note that toward almost all clumps we find bolometric
luminosities in excess of 1000\,L$_{\odot}$ (except clump 8a). This
is particularly surprising for the two starless clump candidates 1 and
3 that exhibit luminosities of 1200 and 1800\,L$_{\odot}$,
respectively, even measured within the smaller apertures of
$\sim$0.5\,pc size.  We will come back to this in section
\ref{luminous}.

\subsection{Kinematic properties from N$_2$H$^+$}
\label{kinematic1}

With the initial single-pointing spectra from \citet{sridharan2005}
one could not associate the different velocity components spatially
well with the sub-clumps of the region. With the new N$_2$H$^+$(1--0)
maps at large scales (Fig.~\ref{n2hplus}) as well as small scales (see
section \ref{pdbi}) we now can study the kinematics of the gas in much
more detail. Figure \ref{n2hplus} presents an overview of the
integrated N$_2$H$^+$(1--0) components around 100 and
50\,km\,s$^{-1}$, respectively. The left panel of Figure \ref{n2hplus}
presents the N$_2$H$^+$ relation with the 70\,$\mu$m warm dust
emission whereas the right panel presents the relation to the 1.2\,mm
cold dust emission.  We find strong N$_2$H$^+$ emission of the
100\,km\,s$^{-1}$ component toward the already very active W43 region
and the younger regions still dark at 70\,$\mu$m without any obvious
preference between them.  In contrast to that, the 50\,km\,s$^{-1}$
component is absent toward the W43 complex but just exhibits emission
associated with three clumps of the IRDC\,18454. Of particular
interest are the two central clumps 1 and 2 which at the spatial
resolution of the Nobeyama data appear to exhibit emission from both
components simultaneously.  As shown in the following section, this is
not exactly true but we can associate the two clumps with
individual velocity components.

Independent of this, the question arises whether these two velocity
components are just chance alignments along the line of site or
whether they are spatially related and may even indicate some
cloud-cloud interaction. Could such a cloud-cloud interaction arise in
the interface region of the Galactic bar and the beginning of the
spiral arm where this region is located? We will come back to this
question in section \ref{converge}.

\begin{figure}[htb]
\includegraphics[angle=-90,width=0.48\textwidth]{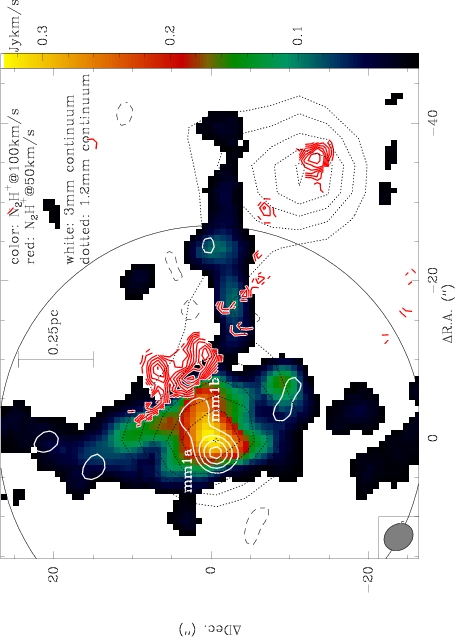}
\caption{PdBI 3\,mm of the clumps 1 and 2. The color-scale shows the
  integrated N$_2$H$^{+}$(1--0) emission within the velocity interval
  from 90.5 to 94.5\,km\,s$^{-1}$ whereas the red contours present the
  N$_2$H$^{+}$(1--0) emission within the velocity interval from 43.8
  to 45.6\,km\,s$^{-1}$. The white contours show the corresponding
  3.2\,mm continuum emission contoured from the 3$\sigma$ level in
  2$\sigma$ steps (1$\sigma=$0.27\,mJy\,beam$^{-1}$). The full circle
  outlines that FWHM of the PdBI primary beam at the given frequency,
  and a scale-bar is shown as well.  The PdBI synthesized beam is
  presented at the bottom-left. The dotted contours show the 1.2\,mm
  continuum emission contoured in 3$\sigma$ levels of
  30\,mJy\,beam$^{-1}$.}
\label{18454-1_100_50}
\end{figure}
\subsection{Zooming into the 2 central clumps}
\label{pdbi}

\subsubsection{Millimeter continuum emission}

\begin{figure*}[htb]
\includegraphics[angle=-90,width=0.99\textwidth]{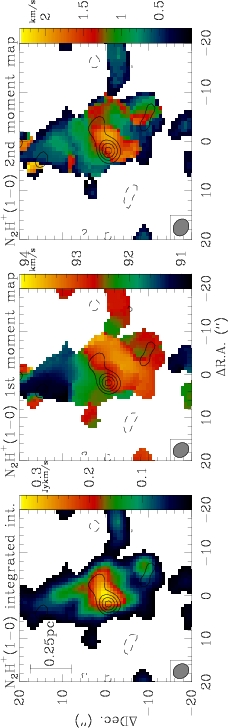}
\caption{The color-scale shows the 0th, 1st and 2nd moment map of the
  N$_2$H$^{+}$(1--0) emission in the velocity regime between 90.5 and
  94.5\,km\,s$^{-1}$ (corresponding to the integrated intensity, the
  intensity-weighted peak velocity and the intensity-weighted
  line-width, respectively). The contours present the 3.2\,mm
  continuum emission contoured from the 3$\sigma$ level in 2$\sigma$
  steps (1$\sigma=$0.27\,mJy\,beam$^{-1}$). The left panel shows a
  scale-bar, and the PdBI synthesized beam is presented at the
  bottom-left of each panel. The 0/0 position is the phase center of
  the PdBI observations at R.A.~(J2000.0) 18:48:02.21 and
  Dec,~(J2000.0) -01:53:55.8.}
\label{18454-1_100_mom}
\end{figure*}

\begin{figure*}[htb]
\includegraphics[angle=-90,width=0.99\textwidth]{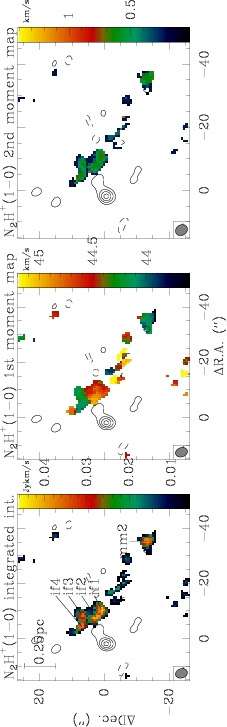}
\caption{The color-scale shows the 0th, 1st and 2nd moment map of the
  N$_2$H$^{+}$(1--0) emission in the velocity regime between 43.8 and
  45.6\,km\,s$^{-1}$ (corresponding to the integrated intensity, the
  intensity-weighted peak velocity and the intensity-weighted
  line-width, respectively). The contours present the 3.2\,mm
  continuum emission contoured from the 3$\sigma$ level in 2$\sigma$
  steps (1$\sigma=$0.27\,mJy\,beam$^{-1}$). The four interface
  positions if1 to if4 and the peak associated with clump 2 are marked
  in the left panel. The left panel shows a scale-bar, and the
  PdBI synthesized beam is presented at the bottom-left of each panel.
  The 0/0 position is the phase center of the PdBI observations at
  R.A.~(J2000.0) 18:48:02.21 and Dec,~(J2000.0) -01:53:55.8.}
\label{18454-1_50_mom}
\end{figure*}

Figure \ref{18454_spitzer} presents a zoom into the central region
covering the two clumps 1 and 2. While both mm emission clumps are
very similar at mm wavelength as well as at 8 and 24\,$\mu$m, clump 2
starts as an emission source from 70\,$\mu$m onwards whereas clump 1
is still dark at these far-infrared wavelengths. This difference is
also manifested in SEDs (Fig.~\ref{seds}) where temperature and
luminosity of clump 1 are lower than the corresponding values for
clump 2. This indicates an earlier evolutionary stage for clump 1.
Figure \ref{18454_spitzer} also shows the FWHM of the primary beam of
our PdBI observations which was centered on clump 1. However, one
should keep in mind that this circle represents only the FWHM, and
that the PdBI can also trace structures outside of that (see below).

The PdBI 3.23\,mm continuum and N$_2$H$^+$ data now allow us a more
detailed characterization of in particular clump 1, but to a lower
level also of clump 2. Figure \ref{18454-1_100_mom} presents the
integrated 100 and 50\,km\,s$^{-1}$ N$_2$H$^+$ emission as well as the
3.23\,mm continuum data. At the higher spatial resolution of $\sim
3.6''$, that corresponds to linear scales of $\sim 20000$\,AU
($\approx 0.1$\,pc), clump 1 resolves into two cores, one being the
dominant and primary component. The projected separation between the 2
cores is $\sim 5.5''$, corresponding to a projected linear separation
of $\sim 30000$\,AU (or $\sim 0.15$\,pc). This is consistent with the
typical Jeans-length of $\sim$0.1\,pc for a gas clump at 20\,K with an
average density of $10^5$\,cm$^{-3}$. The assumed 3.23\,mm continuum
emission from clump 2 is too low to be detected at the edge of the
primary beam of these PdBI observations (Fig.~\ref{18454-1_100_50}).

Peak and integrated fluxes as well as the corresponding gas column
densities and masses (assuming again optically thin dust continuum
emission at a temperature of 20\,K, see section \ref{continuum}) are
presented in Table \ref{3mm}.  The two 3\,mm cores add up to a total
mass of $\sim$91\,M$_{\odot}$, approximately 15\% of the mass detected
in the 1.2\,mm single-dish data (Table \ref{parameters}). This
difference can be attributed to the missing flux typically filtered
out by interferometer observations. As already indicated in section
\ref{continuum}, the interferometrically observed column densities are
larger by about a factor 2 than the single-dish column densities.

\begin{table}[htb]
\caption{Core parameters from the PdBI 3.23\,mm data}
\label{3mm}
\begin{tabular}{lrrrrrr}
\hline \hline
\# & R.A. & Dec. & $S_{\rm{peak}}$ & $S$ & N$_{\rm{H}_2}$ & $M$ \\
   & (J2000.0) & (J2000.0) & $\left(\frac{\rm{mJy}}{\rm{beam}}\right)$ & (mJy) & $\left(\frac{10^{23}}{\rm{cm}^2}\right)$ & (M$_{\odot}$) \\
\hline
1a & 18:48:02.33 & -01:53:56.3 & 0.9 & 1.1 & 3.0 & 59 \\
1b & 18:48:02.01 & -01:53:53.8 & 0.4 & 0.6 & 1.4 & 32 \\
\hline \hline
\end{tabular}
\end{table}

\subsubsection{N$_2$H$^+(1-0)$ spectral line emission}
\label{kinematic2}

The integrated N$_2$H$^+(1-0)$ emission of the 100 and
50\,km\,s$^{-1}$ components observed with the spatial resolution of
the PdBI now for the first time allows us to spatially distinguish the
origin of the different components. Figure \ref{18454-1_100_50} shows
that the 100\,km\,s$^{-1}$ component is closely related to the
starless clump 1. Hence, this starless clump is spatially related to
the western luminous mini-starburst complex W43 (e.g.,
\citealt{nguyen2011}). This is also observable by the filamentary
extension of the 100\,km\,s$^{-1}$ component toward the west, touching
only the northern edge of clump 2.

In comparison to this, the 50\,km\,s$^{-1}$ component is spatially
located at the western edge of clump 1. The projected 50\,km\,s$^{-1}$
component appears to directly touch the western edge of the
100\,km\,s$^{-1}$ component. Technically a bit more surprising, we
find a clear detection of the N$_2$H$^+(1-0)$ 50\,km\,s$^{-1}$
component toward the peak position of clump 2. And at a low level this
50\,km\,s$^{-1}$ clump 2 emission is connected to the 50\,km\,s$^{-1}$
component at the western edge of clump 1.

The visual impression one gets from the close spatial association of
the 50 and 100\,km\,s$^{-1}$ N$_2$H$^+$ data is that these two
velocity components, that are so strongly separated in velocity space,
may still be spatially correlated on the sky and not just chance
projections. It is also intriguing that the 1.2\,mm peak fluxes and
integrated fluxes as well as the projected spatial size of clumps 1
and 2 are so similar (Table \ref{parameters}). While chance projection
of two gas clumps at different distances is obviously a possibility to
explain all features, the whole W43 complex at the interface of the
central bar of our Galaxy and the onset of the spiral arm is known to
exhibit a large breadth of velocity components \citep{nguyen2011}.
Therefore, a spatial association and even interaction between two
complexes could be possible as well (see also section
\ref{kinematic1}). We come back to this scenario in section
\ref{converge}.

Figures \ref{18454-1_100_mom} and \ref{18454-1_50_mom} present the
three moment maps (integrated intensities, intensity-weighted peak
velocities and intensity-weighted line widths) of the 100 and
50\,km\,s$^{-1}$ components, respectively. For the 100\,km\,s$^{-1}$
component we find a velocity gradient approximately
perpendicular to the connecting axis of the two 3.23\,mm peaks mm1a
and mm1b. The broadest line width is also associated with the
strongest mm peak mm1a. The 50\,km\,s$^{-1}$ component does not
exhibit a clear velocity gradient between mm2 and the interface region
of mm1. Furthermore, the 2nd moment map does not show obvious line
width increases toward one or the other position.

\begin{figure}[htb]
\includegraphics[angle=-90,width=0.49\textwidth]{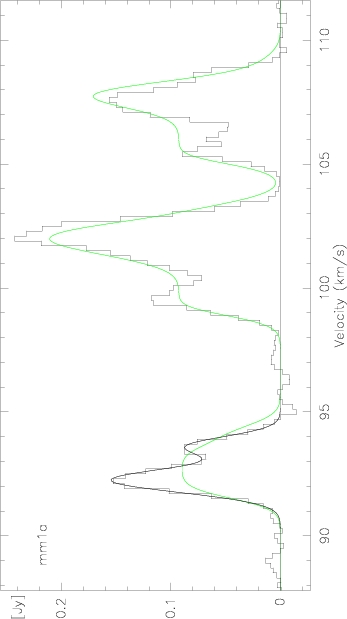}\\
\includegraphics[angle=-90,width=0.49\textwidth]{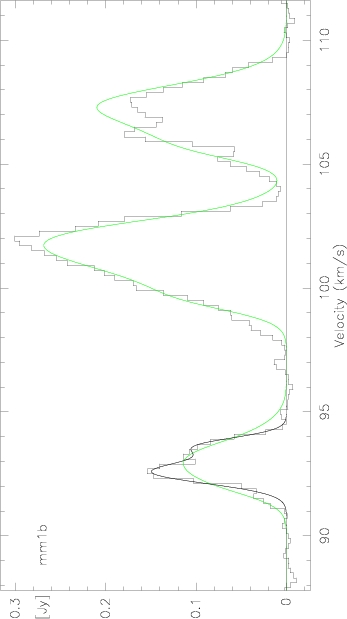}
\caption{N$_2$H$^+(1-0)$ spectra around 100\,km\,s$^{-1}$ toward mm1a
  and mm1b. The green lines show N$_2$H$^+$ fits taking into account
  the whole hyperfine structure whereas the black lines show
  two-component Gaussian fits to the isolated hyperfine structure line
  $-8.0$\,km\,s$^{-1}$ from the central peak. Peak velocities
  (corrected for the $-8.0$\,km\,s$^{-1}$ offset) and line widths of
  the Gaussian fits are presented in Table \ref{n2h+}.}
\label{n2h+_100}
\end{figure}

\begin{figure}[htb]
\centering
\includegraphics[width=0.37\textwidth]{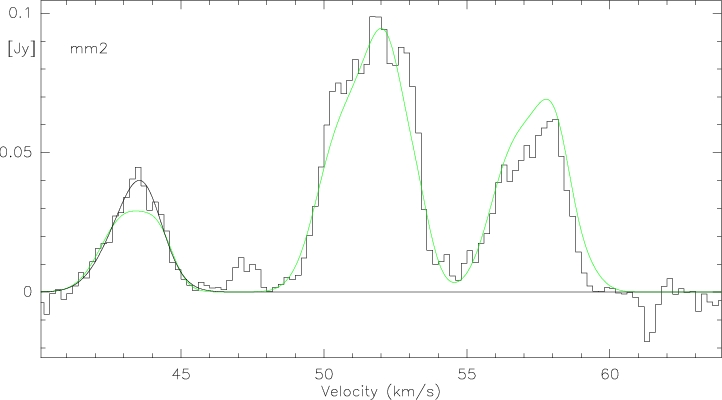}\\
\includegraphics[angle=-90,width=0.37\textwidth]{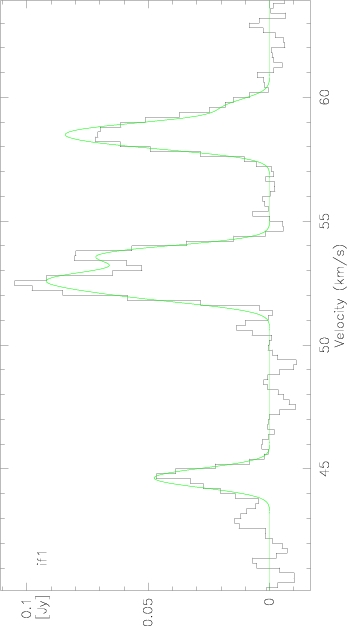}\\
\includegraphics[angle=-90,width=0.37\textwidth]{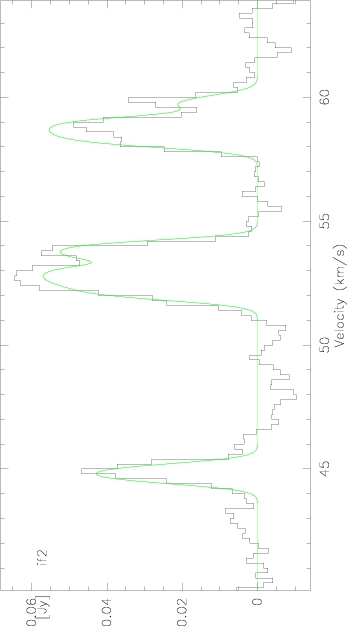}\\
\includegraphics[angle=-90,width=0.37\textwidth]{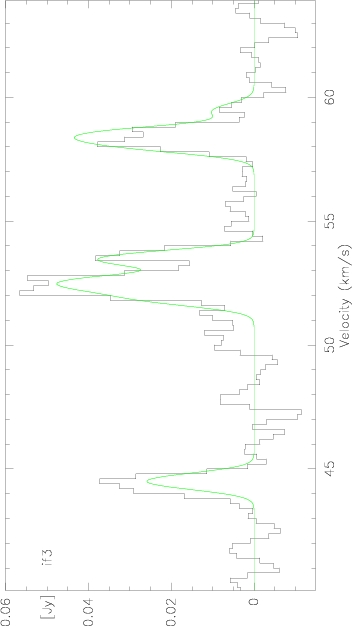}\\
\includegraphics[width=0.37\textwidth]{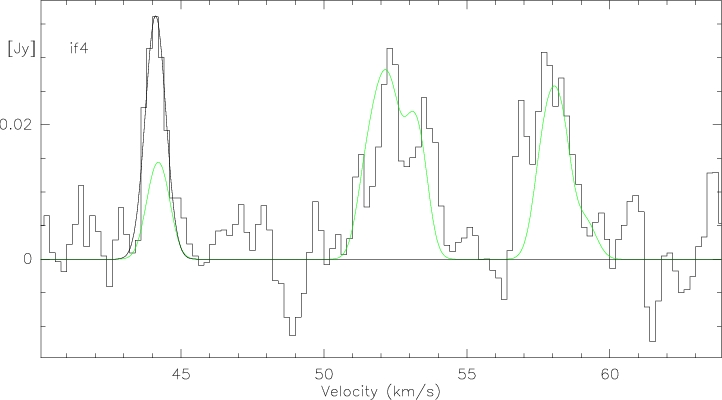}
\caption{N$_2$H$^+(1-0)$ spectra around 50\,km\,s$^{-1}$ toward mm2
  and the four interface positions if1 to if4. The green lines show
  N$_2$H$^+$ fits taking into account the whole hyperfine structure
  whereas the black lines show two- and one-component Gaussian fits to
  the isolated hyperfine structure line $-8.0$\,km\,s$^{-1}$ from the
  central peak. Peak velocities and line widths the hyperfine
  structure fits if1 to if3 and of the Gaussian fits (corrected for
  the $-8.0$\,km\,s$^{-1}$ offset) for mm2 and if4 are presented in
  Table \ref{n2h+}.}
\label{n2h+_50}
\end{figure}

To further investigate the velocity sub-structure of the different
components, we extracted N$_2$H$^+(1-0)$ spectra toward mm1a, mm1b,
mm2 and four positions in the interface region directly west of mm1
(see Fig.~\ref{18454-1_50_mom} for the four selected interface
positions if1 to if4). Figures \ref{n2h+_100} and \ref{n2h+_50} show
the corresponding 100 and 50\,km\,s$^{-1}$ spectra at the different
positions. At the positions of mm1a, mm1b and mm2, the optical depth
is so high that no useful fits to the whole hyperfine structure was
possible. For these positions, we extracted the peak velocities and
line widths from Gaussian fits to the isolated hyperfine structure
component $-8.0$\,km\,s$^{-1}$ offset from the main peak. The spectrum
toward if4 has an unusually strong isolated hyperfine structure
component that the hyperfine fitting was difficult as well. Therefore,
here we also use the fit results from the Gaussian fit to the isolated
hyperfine structure component. For the other three interface positions
if1 to if3, fitting the whole hyperfine structure worked reasonably
well. Table \ref{n2h+} presents the derived peak velocities and line
widths.

\begin{table}[htb] 
\caption{Parameters of N$_2$H$^+$(1--0) emission peaks} 
\begin{tabular}{lrrrrrrrr} \hline \hline
  & $v$    & $\Delta v$ & $M_{\rm{vir}}$ \\
  & km/s   & km/s       & M$_{\odot}$    \\
  \hline
  mm1a & 100.2 & 1.2 & 8.7-13.1 \\
  mm1a & 101.6 & 1.0 & 6.0-9.1  \\
  mm1b & 100.6 & 1.1 & 7.3-11.0 \\
  mm1b & 101.6 & 0.8 & 3.9-5.8  \\
  mm2  & 51.3  & 2.2 & 29.3-44.1\\
  mm2  & 51.8  & 1.5 & 13.6-20.5\\
  if1  & 52.6  & 0.8 & 3.9-5.8  \\
  if2  & 52.8  & 0.7 & 3.0-4.5  \\
  if3  & 52.5  & 0.7 & 3.0-4.5  \\
  if4  & 52.1  & 0.8 & 3.9-5.8  \\
  \hline \hline \end{tabular}
~\\
{\footnotesize For mm1a, mm1b, mm2 and if4, the fits results from the Gaussian fits to the isolated hyperfine structure line (corrected for the $-8.0$\,km\,s$^{-1}$
  offset) are presented. For if1 to if3, the corresponding results from the whole hyperfine structure line fits are listed.}  
\label{n2h+} 
\end{table} 

Independent of being part of two very different general velocity
components at 100 and 50\,km\,s$^{-1}$, the mm peak positions mm1a,
mm1b and mm2 all three show double peaked N$_2$H$^+(1-0)$ emission
where the spectral peaks are separated between 0.5 and
1.4\,km\,s$^{-1}$ (Table \ref{n2h+}). It is interesting to note that
\citet{beuther2009b} found also double-peaked N$_2$H$^+$ spectra
within the IRDC\,19175 region, and recently Ragan et al.~(in prep.)
identified similar signatures in a mid-infrared dark clump of the
Snake IRDC G11.11. Ragan et al.~(in prep.) will discuss this effect in
more detail.

The N$_2$H$^+$ line widths $\Delta v$ on the order of 1\,km\,s$^{-1}$
are larger than the thermal line width at 15\,K of
$\sim$0.15\,km\,s$^{-1}$. However, since our spectral resolution is
only 0.2\,km\,s$^{-1}$ we could not even resolve much narrower lines
properly. Therefore, our measured line width have to be considered as
upper limits, and the actual line widths for some cores may in fact
not be too far off from pure thermal line widths. In contrast to the
luminosities, where we find considerable differences between different
regions (see section luminous for a discussion), the observed
N$_2$H$^+$ line widths of several different IRDCs is very similar
(e.g., IRDC\,18223-3, IRDC\,19175,
\citealt{beuther2005d,beuther2009b}).

Following \citet{maclaren1988}, we calculated the virial masses
$M_{\rm{vir}}=k_2\times R \times \Delta v^2$ with the constant $k_2$
between 190 and 126 for $1/r$ or $1/r^2$ density distributions, the
core radius $R$ (in units of parsecs) and the N$_2$H$^+$(1--0) line
width $\Delta v$. For the radius we assumed all sources to be
unresolved with an average N$_2$H$^+$(1--0) synthesized beam of $\sim
3.6''$, corresponding to the core diameter of 19800\,AU. Table
\ref{n2h+} lists the range of virial masses for the given line widths
and assumed density distributions. Interestingly, the gas masses
derived for mm1a and mm1b from the 3.23\,mm continuum data (Table
\ref{3mm}) are a factor of a few larger than the virial masses
estimated from the N$_2$H$^+$ data. While mass estimates from the dust
continuum emission as well as virial masses both have notoriously large
errors on the order of a few, nevertheless, the data from mm1a and
mm1b are consistent with starless massive gas cores that are likely
gravitationally bound and may collapse and form stars in the near
future.
\section{Discussion}
\label{discussion}

\subsection{Luminous starless clumps}
\label{luminous}

The measured high luminosities from the starless clump candidates
(Fig.\ref{seds}), in particular clumps 1, 3 and 8a are intriguing
since -- down to our detection limits -- they do not have an internal
radiation sources.  Also other sources like clump 2 are interesting
because this region is still dark at 24\,$\mu$m and only becomes an
emission source from 70\,$\mu$m onwards. Hence, it has to be at a
very young evolutionary stage as well. This implies that the
bolometric luminosities of these regions must be largely caused by
external radiation sources of which the prime candidates are the
diverse sources comprising the neighboring mini-starburst W43 (see
Introduction or, e.g., \citealt{motte2003,nguyen2011}). Do we find
differences between IRDCs in the vicinity of very active massive
star-forming regions and more isolated IRDCs?

For comparison, we select the IRDC\,18223 where luminosities and
temperatures were measured in a similar fashion via SED fits of
Herschel data \citep{beuther2010b}. The derived temperatures for the
clumps in IRDC\,18454 are slightly elevated compared to IRDC\,18223
(between 19 and 20\,K for IRDC\,18454, and between 16 and 18\,K in
IRDC\,18223), and the measured bolometric luminosities in IRDC\,18454
are factors between 4 and 10 larger than found in IRDC\,18223. While
the observed aperture size and spatial scale directly translates into
different luminosities (Fig.~\ref{seds}), the smaller aperture of
$19.2''$ (corresponding to $\sim$0.5\,pc) used for IRDC\,18454 is even
smaller than that used for IRDC\,18223 ($\sim$0.64\,pc). Therefore,
spatial scale arguments cannot account for that difference. Another
important parameter for the total bolometric luminosity is the
available gas mass. While clumps 1 and 3 are factors of a few more
massive than IRDC\,18223-s1 and IRDC\,18223-s2, clump 8 presented here
and IRDC\,18223-s2 have approximately the same mass (140 and
170\,M$_{\odot}$, respectively).  Nevertheless, the luminosity of
clump 8 is about a factor 4 larger than that of IRDC\,18223-s2.  It is
no surprise that the temperature differences between IRDC\,18223 and
IRDC\,18454 are less strong than the luminosity differences because
the luminosity has a strong exponential dependence on the temperature
(Stefan-Boltzmann law $L \propto T^4$).

While all these parameters have to be taken a bit cautiously since
the associated errors are large (e.g., for the SEDs the uncertain
background subtraction may contribute significantly), in principle the
errors should be comparable for the different regions because we
applied the same techniques. Therefore, it is tempting to speculate
what may cause the measured luminosity and temperature differences
between infrared dark clumps in the vicinity of the active
star-forming region W43 compared to the more quiescent region
IRDC\,18223. It could be possible that the neighboring mini-starburst
injects considerable energy even into the starless clumps that the
total bolometric luminosities are increased above their expected more
quiescent values. To first order, with such a luminosity increase one
also expects a temperature increase within the gas clumps. Although
the measured temperature difference between the regions is less strong
than the luminosity increase, nevertheless, we see elevated
temperatures in IRDC\,18454 compared to IRDC\,18223. Considering the
large apertures used for the SEDs, the derived average temperatures
may still be dominated by the average temperature of the general ISM
that is in the same range (e.g., \citealt{reach1995}).

We can also quantitatively estimate the influence of the total
bolometric luminosity of the Wolf-Rayet/OB cluster ($\sim 3.5\times
10^6$\,L$_{\odot}$) on the gas and dust environment at the typical
projected separations from the target IRDCs (Fig.~\ref{spire_pacs}).
Assuming that all radiation of the central cluster is reprocessed as
far-infrared radiation, and using an approximate projected separation
of the infrared dark clumps of 11\,pc, as well as Gaussian source FWHM
of $36.6''$ or $19.2''$, respectively (Sec.~\ref{sedsection} and
Fig.~\ref{seds}), the received luminosity at the location of the IRDCs
are on the order of 2500 and 700\,L$_{\odot}$, respectively.
Comparing these values with the measured luminosities in the same
given apertures (Fig.~\ref{seds}), elevated luminosities caused by the
nearby luminous cluster appear clearly feasible. Furthermore, one can
estimate the approximate required interstallar radiation field to
account for the observed luminosity. A clump with a surface
corresponding to a radius $19.2''/2$ has to intercept an interstellar
radiation field of $\sim$306 (in units of the Habing field
$G_0=1.6\times 10^{-3}$\,erg\,s$^{-1}$cm$^{-2}$, \citealt{habing1968})
to exhibit a luminoisty of 1000\,L$_{\odot}$.  In comparison, if one
just accounts for geometric dillution, the $\sim 3.5\times
10^6$\,L$_{\odot}$ of the Wolf-Rayet/OB cluster correspond at a
distance of 11\,pc to $\sim$588 times the interstellar radiation
field. Subtracting additional scattering and blockade by interfering
gas clumps, these two values agree well with each other. Based on
kinematic arguments, \citet{motte2003} also discussed the influence of
the luminous cluster on the environment, and they similarly infer that
there should be considerable influence from that first generation of
high-mass star formation on its environment.

\begin{figure}[htb]
\includegraphics[width=0.49\textwidth]{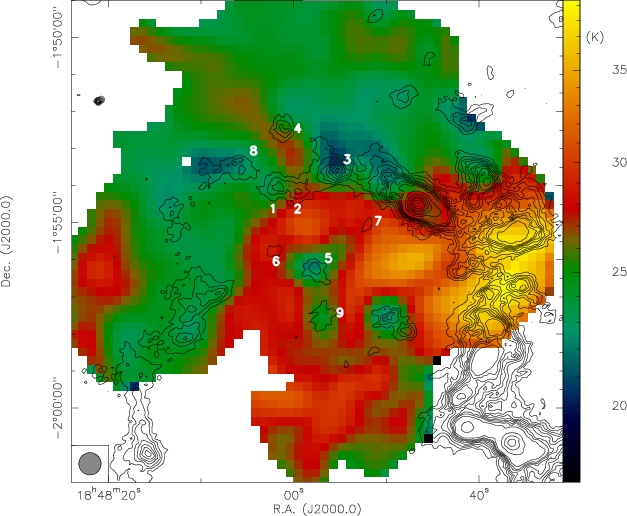}
\caption{The color-scale shows a temperature map in K derived for the
  whole region from the combined Herschel and ATLASGAL data between 70
  and 870\,$\mu$m, all smoothed to the spatial resolution of the
  500\,$\mu$m data. Contour levels of the overlaid 870\,$\mu$m data
  start at the 3$\sigma$ levels of 0.27\,mJy\,beam$^{-1}$ and continue
  in in 3$\sigma$ steps to 2.7\,Jy\,beam$^{-1}$, from where they
  continue in 2.7\,Jy\,beam$^{-1}$ steps. Our target clumps are
  marked.}
\label{t_herschel}
\end{figure}

In  addition  to  the   individual   SED fits   presented  in  section
\ref{sedsection} and Figure \ref{seds},  one can also smooth  all maps
between 70  and  870\,$\mu$m to  the  same  spatial  resolution of the
500\,$\mu$m   data  and   fit   the SEDs pixel  by    pixel  (see also
\citealt{stutz2010} for  details  of the  fitting).  This  large-scale
fitting approach allows us to infer the dust temperature structure and
gradients across the whole field. Figure \ref{t_herschel} presents the
results.  As expected, we identify  a clear temperature gradient  from
more than 35\,K toward  the W43 main  complex down to low values below
20\,K in the east  of the region. In  particular, we see that  many of
the  submm continuum   structures   correlate  well  with the   lowest
temperatures.  This is true  for the dark   clumps of this analysis as
well as for the filament  going approximately in northsouth  direction
at the eastern edge of  the map. The large-scale east-west temperature
gradient throughout   the   region is  consistent   with   significant
influence of the   W43 complex  on  the here  discussed infrared  dark
high-mass starless clump candidates.

Since the Jeans length and Jeans mass as indicators of the
fragmentation properties of star-forming gas clumps are proportional
to the temperature ($\lambda_J \propto T^{1/2}$ and $M_J \propto
M^{3/2}$), increases in luminosity and temperature of young gas clumps
could change their fragmentation properties to preferentially form
larger and more massive fragments. Although still speculative, such a
scenario could imply that nearby active star formation does not only
have a destructive character destroying pristine gas clumps, but that
the energy input of close-by active regions could even elevate the
ability to form massive stars in a triggered fashion in the immediate
environment of the earlier generations of stars. While the energy
input from a first generation of low-mass stars appears to be too
little to halt fragmentation \citep{longmore2011}, the influence of
the nearby luminous high-mass star-forming region may be more
important.

\subsection{Colliding clouds or chance alignment?}
\label{converge}

As indicated in sections \ref{kinematic1} \& \ref{kinematic2}, the two
different velocity components at 100 and 50\,km\,s$^{-1}$ may either
be chance projections along the line of sight, or they could be
associated and potentially interacting cloud complexes. Based mainly
on large-scale $^{13}$CO(1--0) data from the Galactic Ring survey
(GRS, \citealt{jackson2006}), \citet{nguyen2011} suggest that the
velocities associated directly with W43 range between 60 and
120\,km\,s$^{-1}$, whereas lower velocity components are unlikely part
of the W43 complex.

\begin{figure}[htb]
\includegraphics[width=0.49\textwidth]{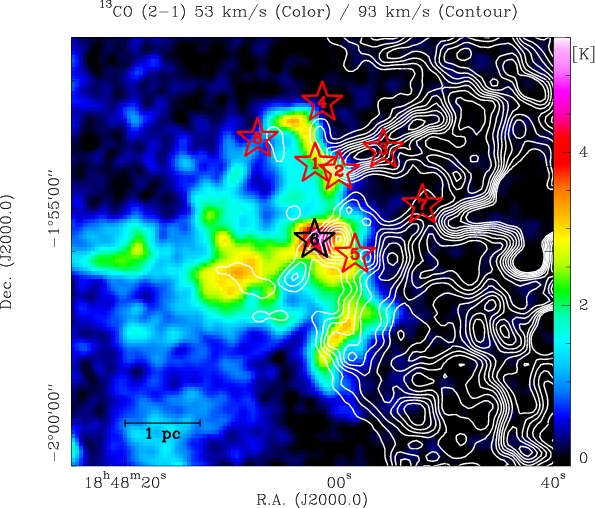}
\caption{IRAM\,30m $^{13}$CO(2--1) observations toward a larger part
  of W43 and IRDC\,18454. The color-scale shows the $^{13}$CO emission
  around 53\,km\,s$^{-1}$ whereas the contours present gas at
  93\,km\,s$^{-1}$. The stars mark the positions of our
  clumps 1 to 8, a scale-bar is shown at the bottom-left.}
\label{13co}
\end{figure}

In contrast to this, the spatial structure of the two velocity
components as presented in Figs.~\ref{n2hplus} and
\ref{18454-1_100_50} indicates that the two components may also be
spatially associated and interacting. Also the fact that the mm fluxes
of clump 1 and 2 are so similar could be taken as circumstantial
support for this idea. 

To further investigate whether we find additional signatures of
associated clouds in that region, we investigated the $^{13}$CO(2--1)
data of a recent IRAM 30\,m large program to study the entire W43
complex (PI F.~Motte). The whole project, the data and more detailed
analysis will be presented in forthcoming papers by Nguyen-Luong et
al.~and Carlhoff et al., here we only show a small sub-set where two
selected velocity components are compared. Figure \ref{13co} presents
an overlay of the $^{13}$CO(2--1) emission around 50 and
100\,km\,s$^{-1}$ over a larger area but also encompassing our region
of interest. The two velocity components appear in projection as a
layered structure where one velocity component mainly emits at
locations that are devoid of emission from the other velocity
component. Unfortunately, this is also no unambiguous proof for
associated or interacting clouds.  However, the spatial alignment of
different velocity components for molecules with varying critical
densities over a broad range of spatial scales is suggestive for the
idea of connected and interacting structures.

Additional interesting information about different velocity components
in this region is found for other tracers and wavelengths as well. For
example, CH$_3$OH Class {\sc ii} maser were found in this region at
both velocities (e.g., \citealt{menten1991}). Furthermore, the recent
H{\sc ii} region radio recombination line survey of
\citet{anderson2011} detected toward 21 out of 23 targeted position
within 0.5 degrees from W43 two velocity components as well (one of
the positions is marked in Fig.~\ref{spire_pacs}). While projection
effects of molecular and atomic gas along the line of sight through
the Galaxy are well known, having such a large number of
multiple-velocity recombination line spectra is more surprising since
one does not expect that many projected H{\sc ii} regions along each
line of sight. Hence interacting clouds and H{\sc ii} regions are
possible. Furthermore, it is interesting to note that the largest
number of H{\sc ii} regions in our Galaxy are also found toward
Galactic longitudes around $l\approx \pm30$\,degrees
\citep{wilson1970,lockman1979,anderson2011}, hence again including the
W43/IRDC\,18454 complex. All this evidence suggests special
environmental physical properties in this region that may affect the
star formation processes.

How could such interacting velocity components be explained? The whole
W43-IRDC\,18454 complex is located at approximately a Galactic
longitude of 30.85 degrees where the Galactic bar ends and the Scutum
spiral arm starts (for a recent description of the different Galactic
components, see e.g., \citealt{churchwell2009,nguyen2011}). Hence, one
could fathom interaction between Galactic bar and spiral arm
structures, e.g., streaming motions between both Galactic components.
Following the Galaxy modeling by \citet{bissantz2003}, the inner
spiral arms connecting to the bar and the bar should co-rotate.  In
this framework, the 100\,km\,s$^{-1}$ component should be associated
with the beginning of the Scutum arm (see also \citealt{vallee2008}).
Spiral arm velocities for the Sagittarius and Perseus arm are,
according to \citet{vallee2008}, around 70 and 25\,km\,s$^{-1}$,
respectively.  However, there is no clear association with a
50\,km\,s$^{-1}$ gas component. Inspecting the Galactic plane CO
survey by \citet{dame2001}, at a Galactic longitude of $\sim 30.85$
degrees one also find strong emission at 100\,km\,s$^{-1}$ and only
less prominent features at 50\,km\,s$^{-1}$. Therefore, from the
spiral arm structure of our Galaxy, it is not obvious to what distance
one can associate the 50\,km\,s$^{-1}$ component. Furthermore,
\citet{nguyen2011} associate even 60\,km\,s$^{-1}$ gas with W43 and
not so much with the Sagittarius arm as one would expect from the
Galaxy models (e.g., \citealt{vallee2008}). Although we cannot firmly
establish whether the 100 and 50\,km\,s$^{-1}$ gas components are
spatially associated or just chance alignments, the close projected
association of them as well as the extremely busy and active nature of
that part of the Galaxy at the interface of the Galactic bar with the
inner spiral arm, where streaming motions between both components are
likely to occur, makes spatial association and interaction of
different cloud components here a fair possibility. See also
\citet{anderson2011} for a similar discussion of the origin of the
ionized gas.

One way to distinguish the different explanation is to derive accurate
distances to the different components. And the best way to do that
today is parallax measurements of maser sources (e.g.,
\citealt{reid2009}). The VLBI now has a very large project to do that
for many maser sources throughout the Galactic plane (PI M.~Reid), and
the masers toward the W43 complex are part of that program
(K.~Menten, priv.~comm.). Therefore, we expect to get accurate
distances for the different velocity components in the coming years.
That should then allow us to answer the question whether these
different velocity components are indeed spatially associated or
whether they are just chance projections along the lines of sight.


\section{Conclusions}
\label{conclusion}

The W43 complex is know as one of the most active sites of star
formation in our Galaxy. Our aim in that context was to investigate
the nature of very young massive gas clumps in the vicinity of W43
prior to star formation or at very early evolutionary stages. A
multi-wavelength continuum study of the region between 70\,$\mu$m and
1.2\,mm wavelengths allows us to isolate gas clumps that are still
dark at 70\,$\mu$m wavelengths. These are among the best candidates of
being genuine high-mass starless cores. An analysis of the spectral
energy distributions finds temperatures between 19 and 21\,K and
luminosities usually larger than 1000\,L$_{\odot}$ for these high-mass
starless clump candidates. These values are larger than those found
toward infrared dark clouds in more quiescent environments. A clear
temperature gradient from the luminous W43 complex to the studied
starless clump candidates is identified. We discuss whether the nearby
mini-starburst W43 not only has destructive effects on nearby gas
clumps, but that elevating the average gas temperatures can increase
the Jeans length and Jeans mass, hence favoring the formation of
massive stars in the environment.

An analysis of the spectral data obtained with single-dish instruments
and the Plateau de Bure interferometer reveals two different velocity
components at 100 and 50\,km\,s$^{-1}$. While chance projection along
the line of sight is possible, the projected spatial association of
the two gas components indicates that the two components may be
spatially associated or even interacting. We discuss the general
velocities expected in that part of our Galaxy. While the
100\,km\,s$^{-1}$ component associated also with W43 appears to be
clearly part of the Scutum arm that starts at the end of the inner
Galactic bar, it is more difficult to clearly associate the
50\,km\,s$^{-1}$ with a distinct Galactic feature.  Therefore, it
remains possible that the extremely active location of that region at
the interface of the Galactic bar and the inner spiral arm hosts
clouds at different velocities that may be interacting.



\end{document}